# Efficient and accurate determination of lattice-vacancy diffusion coefficients via nonequilibrium *ab initio* molecular dynamics

D.G. Sangiovanni,[1,*] O. Hellman,[1] B. Alling,[1,2] and I.A. Abrikosov[1,3,4]

[1]Department of Physics, Chemistry and Biology (IFM)
Linköping University, SE-58183 Linköping, Sweden
[2]Max-Planck-Institut für Eisenforschung GmbH, D-402 37 Düsseldorf, Germany
[3]Materials Modeling and Development Laboratory, National University of Science and Technology 'MISIS', 119049 Moscow, Russia
[4]LACOMAS laboratory, Tomsk State University, 634050 Tomsk, Russia

We revisit the color-diffusion algorithm [P. C. Aeberhard *et al.*, Phys. Rev. Lett. **108**, 095901 (2012)] in nonequilibrium *ab initio* molecular dynamics (NE-AIMD), and propose a simple efficient approach for the estimation of monovacancy jump rates in crystalline solids at temperatures well below melting. Color-diffusion applied to monovacancy migration entails that one lattice atom (colored-atom) is accelerated toward the neighboring defect-site by an external constant force **F**. Considering bcc molybdenum between 1000 and 2800 K as a model system, NE-AIMD results show that the colored-atom jump rate $k_{NE}$ increases exponentially with the force intensity F, up to F values far beyond the linear-fitting regime employed previously. Using a simple model, we derive an analytical expression which reproduces the observed $k_{NE}(F)$ dependence on F. Equilibrium rates extrapolated by NE-AIMD results are in excellent agreement with those of unconstrained dynamics. The gain in computational efficiency achieved with our approach increases rapidly with decreasing temperatures, and reaches a factor of four orders of magnitude at the lowest temperature considered in the present study.

*Corresponding author: Davide G. Sangiovanni, Tel. 0046 13282623, Fax 004613137568,
e-mail: davsan@ifm.liu.se

# I. Introduction

Point defects, such as lattice vacancies, affect the chemical, structural, mechanical, optical, and electronic properties of crystalline solids.[1, 2] Transition-state theory (TST),[3] routinely used to extrapolate finite-temperature vacancy jump rates from migration energies and attempt frequencies calculated at 0 K,[4] provides reliable point-defect diffusivities for temperatures at which the lattice vibrations remain essentially harmonic.[5] However, the modifications induced by the collective atomic motion on the effective potential energy landscape, which in turn controls atomic mobilities, become progressively more significant for increasing temperatures.[6-8] Moreover, diffusion parameters calculated at 0 K are not reliable for crystal structures which are stable at finite temperature due to anharmonic lattice vibrations, while exhibit imaginary phonon frequencies at 0 K.[9, 10] This results in molecular dynamics (MD) becoming the primary computational tool for evaluating reaction rates and revealing the occurrence of non-intuitive atomistic processes at finite temperatures.[11, 12]

Accurate density-functional-based *ab initio* MD (AIMD) is limited, due to high computational cost, to the modeling of small systems (few hundreds atoms) during relatively short time frames (few ns). Determining the rate of rare events via AIMD is a challenging task. Thus, one needs to resort to methods which can selectively accelerate a physical process of interest. The use of fictitious driving forces combined with a synthetic thermostat, which dissipates the energy increase due to external work, allows to accurately model a nonequilibrium steady state via MD.[12] As long as the system's response to the applied constraint is linear, equilibrium (unconstrained) rates are reliably extrapolated from the results of nonequilibrium MD.[13, 14] However, restricting the acceleration method to linear-fitting ranges provides satisfactory gain in computational efficiency only for studying migration processes occurring with relatively high frequencies. While, in general, deriving an analytical expression of the nonequilibrium rate dependence on the intensity of external driving forces beyond the linear jump-rate vs. force trend is a complex problem,[12] in cases for which this dependence can be accurately modeled, equilibrium rates can be efficiently retrieved from nonlinear-range results.

With the color-diffusion (CD) algorithm,[14, 16] an extension of the tagged-particle method,[17] an external constant force-field **F** is used to speed-up migration of an atomic species, while coupling all other particles to an isokinetic[18] thermostat. Recently, CD has been employed in nonequilibrium AIMD (NE-AIMD) investigations of Li$^+$ migration in solid LiBH$_4$ [14], in which equilibrium diffusion coefficients are obtained by fitting linear-range results. However, tasks as, e.g., the estimation of vacancy diffusivities in crystalline solids remain unfeasibly time consuming for NE-AIMD restricted to the CD linear-fitting regime.

In this work, we show that, for cases in which atomic migration takes place over a single energy barrier (assumption typically valid for point-defect diffusion in crystal lattices[19-22] and adatom dynamics on surfaces[23,24]), equilibrium jump rates can be efficiently retrieved by fitting accelerated rates obtained within a sharply nonlinear jump-rate vs. force trend. To this aim, the CD algorithm is applied in NE-AIMD simulations of monovacancy jump in bcc Mo, here chosen as model system. Numerical NE-AIMD experiments show that nonequilibrium (accelerated) rates $k_{NE}$ follow an Arrhenius-like behavior up to intensities F far beyond the linear range. We propose an analytical description for $k_{NE}(F)$ which, at the limit of vanishing F, closely approaches non-accelerated equilibrium rate ($k_E$) values. The extension of the CD method to the exponential regime translates, at simulation temperatures much lower than the melting point, in computational times several orders of magnitude smaller than those required in non-accelerated dynamics.

**II. Computational details and methods**

AIMD runs are performed with VASP [25] using the projector augmented wave method[26] and the Armiento-Mattsson functional (AM05).[27] The equations of motion are integrated at 1 fs time intervals within the canonical ensemble NVT. At each time step, the total energy is evaluated to an accuracy of 10$^{-5}$ eV/atom using a plane-wave energy cutoff of 400 eV, while sampling the Brillouin zone Γ point. The electronic thermal excitations are modeled with electron-smearing energies corresponding to $k_BT$. In non-accelerated simulations, prior to counting diffusion events, the lattice vibrations are allowed to stabilize for four ps by controlling the temperature via the Nosé-Hoover thermostat. The AM05 bcc Mo lattice parameter is only ~0.5% smaller than the room temperature experimental value (3.147 Å). This allows us to use experimental equilibrium volumes[28, 29] for each simulation temperature. Simulation box sizes, 127 Mo atoms (*n*) and one vacancy (4x4x4 bcc unit cells), are sufficient to avoid point-defect self-interactions.[30] During both equilibrium and non-equilibrium simulations, vacancy jumps are identified from the variations in atomic positions.[31]

The melting point $T_{melt}$ of bcc Mo is ~2900 K.[29] Simulation temperatures close to $T_{melt}$ allow us to obtain well-converged AIMD vacancy jump rates $k_E$, defined as lattice-atom hops per time unit, which can be used to assess the accuracy of equilibrium rates $k_{NE \to E}$ extrapolated by NE-AIMD results. $k_E$ is evaluated at temperatures T = 1600, 1800, 2000, 2400, 2600, and 2800 K with simulation times of approximately 1.1, 0.7, 0.5, 0.1, 0.1, and 0.1 ns, respectively. Jump activation energies $E_a$ and attempt frequencies A are obtained by fitting $\ln[k(T)]$ vs. $1/T$ using the Arrhenius expression $k(T) = A \cdot \exp[-E_a/(k_BT)]$. Uncertainties on k, A, and $E_a$ values are determined as described in reference.[12]

Equilibrium rates calculated in this work [$k_E$(2800 K)=6.14(x1.2$^{\pm1}$)x10$^{11}$ s$^{-1}$, $k_E$(2600 K)= 4.94(x1.2$^{\pm1}$)x10$^{11}$ s$^{-1}$] compare well with those of previous AIMD estimations [$k_E$(2800 K)= 6.5(x2.0$^{\pm1}$)x10$^{11}$ s$^{-1}$, $k_E$(2600 K)=3.0(x2.0$^{\pm1}$)x10$^{11}$ s$^{-1}$].[30] The linear fit of $\ln[k_E(T)]$ vs. $1/T$ yields A=44.9(x3.0$^{\pm1}$)x10$^{13}$ s$^{-1}$ and $E_a$=1.58±0.25 eV, in excellent agreement with the experimental value $E_a$=1.62±0.27 eV.[32] All Mo jumps recorded in AIMD runs occur among nearest-neighbor lattice sites; primary diffusion mechanism in bcc metals.[33] Thus, in this work, we consider only <111> vacancy migration.[34]

Color-diffusion applied to monovacancy migration in crystal lattices can be summarized as follows. Immersing a thermally-equilibrated system in a constant force-field **F**, the force acting on atom *j* is $c_j \cdot \mathbf{F}$, for which $c_j$ (colors) are scalars. One lattice atom, referred to as *colored-atom* ($c_1$ =

1), is pushed by a force **F** toward the neighboring vacancy, while balancing forces -**F**/(*n*-1) act on all other atoms [$c_{2,...,n}=-(n-1)^{-1}$]. The system temperature is controlled via rescaling all but the colored-atom velocities at each time step.

For each force-field intensity F used at a temperature T, nineteen initial NE-AIMD configurations are randomly selected out of seventy equilibrated states, in turn extracted every ps from AIMD runs. We use f values ($f = F/\sqrt{3}$ is the magnitude of <100> **F** components) in the ranges 0.6–1.1 eV/Å with T = 1000, 1200, and 1400 K, 0.3–1.1 eV/Å with T = 1600 and 1800 K, and 0.1–1.6 eV/Å with T = 2000 K, at f intervals of 0.1 eV/Å. Each NE-AIMD simulation is terminated when a neighboring Mo atom has migrated into the vacant site or when the simulation wall-time has been reached. $k_{NE}$ is obtained by dividing the number (over all sample runs) of colored-atom jumps (multiplied by the number of nearest neighbors, 8, in bcc unit cells) by the total simulated time.

## III. Results

Fig. 1 shows that the accelerated Mo vacancy jump rate $k_{NE}(F)$ calculated at 2000 K, normalized by the equilibrium $k_E$(2000 K) value, follows an exponential-like dependence on F up to $F \cong 1.8$ eV/Å. NE-AIMD results at 1800 and 1600 K exhibit the same behavior. To clarify the effect of an applied force on the colored-atom jump rate, we use a simple TST-based one-dimensional model. Within the model, the zero-field effective potential energy landscape $E_{LS0}$ probed by the colored-atom along a straight migration path **x** [34] is approximated by a sinusoidal function $E_{LS0}(x) = E_{a0}\left[-\cos(\pi \cdot x / x_{TS0}) + 1\right]/2$. Note that equilibrium (unperturbed) model-quantities are indicated by subscripts ending with 0. $E_{LS0}(x)$ is characterized by two zero-energy minima: at the colored-atom initial (x = $x_{eq0}$ = 0) and final (x = vacancy-site $x_{vac}$) equilibrium positions, and one maximum of energy $E_{a0}$ at the transition state, x = $x_{TS0}$, mid-point of the diffusion-path (Fig. 2).

The effect of an external force **F** (aligned with **x**) on the perturbed potential energy profile $E_{LS}(x)$ *seen* by the colored-atom is interpreted as follows. After a short transient period, $E_{LS}(x)$ reaches a steady-state $E_{LS}(x, F) = E_{LS0}(x) - F \cdot x$. For increasing intensities F, the colored-atom equilibrium position $x_{eq}$ and the transition state position $x_{TS}$ shift continuously toward each other (Fig. 2). At the same time, the effective jump activation energy, $E_a(F) = E_{LS}[x_{TS}(F), F] - E_{LS}[x_{eq}(F), F]$, becomes progressively smaller. The jump attempt frequency, approximated within the harmonic limit as $A(F) \propto \sqrt{\left[\partial^2 E_{LS}(x, F)/\partial^2 x\right]_{x_{eq}(F)}}$, decreases monotonically with increasing F due to the change in $x_{eq}(F)$. When F reaches the value ($F_{max}$) for which $x_{eq}$ and $x_{TS}$ meet at the sinusoidal-curve flex-point, the harmonic approximation is no longer valid, and the model becomes dynamically unstable.

Starting from the unperturbed $E_{LS0}(x)$ sinusoidal shape, we numerically evaluate the model accelerated-rate k dependence on F, $k(F) \propto A(F) \cdot \exp[-E_a(F)]$, for F varying up to $F_{max}$ (Fig. 3). The model k(F) vs. F trends illustrated in Fig. 2 are consistent with those of NE-AIMD $k_{NE}(F)$ values up to $F \cong 0.9\, F_{max}$ (compare Fig. 3 with Fig. 1). Since the acceleration factor $k_{NE}(F)/k_E$ increases rapidly with F, Fig. 1, the aim is to describe A(F) and $E_a(F)$ up to F close to $F_{max}$.

Our analytical derivation of A(F) and $E_a(F)$ based on the TST model can be found in [31]. In synthesis, fitting the model k dependence on F with:

$$k(F, T) = k_0(T) \cdot \exp\left[\frac{x_{TS0}(T)}{k_B T} \cdot F - \alpha(T) \cdot F^2\right], \quad (1)$$

where $\alpha(T)$ is a temperature-dependent parameter and $k_0$ is the equilibrium jump rate, reproduces the $k_{NE}(F)$ trends up to $F' \cong 0.75 \cdot F_{max}$ (Fig. 3). Taking the logarithm of both sides of Eq. (1) yields a polynomial $y_T(F)$:

$$\ln[k(F, T)] = \ln[k_0(T)] + \frac{x_{TS0}(T)}{k_B T} \cdot F - \alpha(T) \cdot F^2. \quad (2)$$

For fields of vanishing intensities, the slope of the parabola becomes $\lim_{F\to 0}\frac{\partial[y_T(F)]}{\partial F}=\frac{x_{TS0}(T)}{k_B T}$.

Setting $x_{TS0}(T)=d_{NN}(T)/2$ (for which $d_{NN}(T)$, equilibrium nearest-neighbor distance at a temperature T, can be obtained, accounting for the thermal expansion, by rescaling the 0 K $d_{NN}$ value), the unknown parameters remaining in Eq. (2) are $\alpha(T)$ and the quantity of practical interest, the equilibrium jump rate $\ln[k_0(T)]$.

Let us return to *ab initio* simulations of vacancy migration in Mo. NE-AIMD jump rate $k_{NE}(F,T)$ dependences on F plotted in Fig. 4 for T = 1600, 1800, and 2000 K, are consistent with those seen in our model calculations (Fig. 3). The results obtained at 1000, 1200, and 1400 K exhibit the same behavior. Although the actual 0-K energy profile along the reaction coordinate deviates from sinusoidal,[34, 35] the expression in Eq. (2), based on a sine function model-Hamiltonian, accurately reproduces the $k_{NE}(F)$ behavior up to F′ at all temperatures.

At 2000 K, $x_{TS0} \cong 1.38$ Å. Using $E_{a0}$ determined by AIMD, 1.58±0.25 eV, in the expression $F_{max}=\frac{\pi \cdot E_{a0}}{2\,x_{TS0}}$ [31] yields $F_{max}$=1.80±0.25 eV/Å; consistent with the force interval corresponding to zero slope in the $k_{NE}(F)$ curve of Fig. 1. If neither $x_{TS0}$ nor $E_{a0}$ vary significantly within the probed temperature range, $F_{max}$ can be estimated by NE-AIMD at a given temperature, and then assumed to remain constant. Thus, F′ $\cong$ 1.4 eV/Å (<100> components f′ = 0.8 eV/Å) is the maximum intensity used in the fit (Eq. 2) of NE-AIMD results. The equilibrium rates ($k_{NE\to E}$) extrapolated from $k_{NE}(f)$ obtained within the range f=0.1–0.8 eV/Å, are in excellent agreement with AIMD $k_E$ values (inset in Fig. 4). This validates our approach and lends confidence that equilibrium rates can be reliably obtained by fitting nonequilibrium results at all temperatures. It is worth stressing that the acceleration factor $k_{NE}/k_E$ achieved in our approach increases rapidly for decreasing temperatures (Fig. 4); restricting CD, as done in previous investigations,[14, 15] to the narrow F interval yielding a *quasi*-linear $k_{NE}(F)$ trend would provide considerably lower gain in computational efficiency (inset in Fig. 1).

Extrapolated $k_{NE\to E}$ values progressively approach equilibrium AIMD rates $k_E$ with increasing number of $k_{NE}(f)$ interpolation points within the f range 0.8–0.3 eV/Å (Fig. 5). In fact, $k_{NE\to E}$ obtained by fitting just three $k_{NE}(f)$ points (f = 0.6, 0.7, and 0.8 eV/Å) falls within, or at least close to, AIMD uncertainty intervals. Thus, f = 0.6, 0.7, and 0.8 eV/Å is used to assess equilibrium rates at 1400, 1200, and 1000 K: $k_{NE\to E}$ = 8.2x10$^8$, 4.3x10$^7$, and 1.5x10$^6$ s$^{-1}$, respectively. At these temperatures, the estimation of vacancy jump rates via non-accelerated dynamics would require simulation times of several µs. However, the use of the CD algorithm within the exponential regime reduces simulation times to a total of ~2 ns.

Finally, we compare the results of our approach with Mo vacancy diffusion coefficients D(T) determined experimentally between 1350 and 2800 K.[36] Since $D(T) \propto k(T) \cdot c_V(T)$, theoretical prediction of D(T) requires the evaluation of vacancy equilibrium concentrations $c_V(T)$.

Glensk et al.,[7] using *ab initio* calculations accounting for anharmonic effects, have shown that the temperature-dependence of the Gibbs vacancy formation energy G$^f$ in fcc Al and Cu is accurately described by $G^f(T) = H^f_{0K} - \frac{1}{2}T^2 S'$, where $S' = \frac{\partial S^f}{\partial T}$ is a constant as the formation entropy, $S^f(T) \cong T \cdot S'$, increases approximately linearly with T. The formation enthalpy $H^f_{0K}$ of Mo vacancies estimated *ab initio*, 3.1 eV [30], is close to the experimental value 3.24±0.09 eV.[32] S' can be obtained from S$^f$ measured at ~620 K [32] as: S' = [(2.6±0.4) k$_B$]/(620 K). Thus, using theoretical k(T) values in the expression $D(T) = \frac{1}{6}\gamma \cdot d^2_{NN}(T) \cdot c_V(T) \cdot k(T)$, for which $c_V(T) = \exp[-G^f(T)/(k_B T)]$ and γ=0.72 is the correlation factor for vacancy jump in bcc crystals,[37] we determine D(T) for T between 1000 and 2800 K. Calculated D(T) values agree well with experimental results[36] at all temperatures (Fig. 6).

**IV. Conclusions**

In summary, we show that the CD algorithm, employed in NE-AIMD simulations of Mo monovacancy diffusion accelerated by force fields reaching intensities far beyond the linear k vs. F regime, allows to reduce computational times by several orders of magnitude. The method is validated by comparing equilibrium jump rates extrapolated from the results of accelerated dynamics with non-accelerated AIMD jump rates. The accuracy of extrapolated equilibrium rates can be improved systematically by increasing the number of NE-AIMD interpolation points, yet maintaining considerably shorter computational times compared to those required in AIMD simulations. Our theoretical estimations of Mo vacancy diffusivities are in agreement with experimental results.

**Acknowledgments**

The calculations and simulations were performed, with resources provided by the Swedish National Infrastructure for Computing (SNIC), on the Gamma, Triolith, Matter, and Kappa Clusters located at the National Supercomputer Centre (NSC) in Linköping and on the Beskow cluster located at the Center for High Performance Computing (PDC) in Stockholm, Sweden. Dr. K. Finzel is acknowledged for useful discussions. W. Olovsson and P. Münger at NSC, and H. Leskelä and J. Vincent at PDC are acknowledged for assistance with technical aspects. We gratefully acknowledge financial support from the Knut and Alice Wallenberg Foundation (Isotope Project No. 2011.0094), the Swedish Research Council (VR), B. Alling grants: 621-2011-4417 and 330-2014-6336, Linköping Linnaeus Initiative LiLi-NFM (Grant No. 2008-6572), and the Swedish Government Strategic Research Area Grant in Materials Science on Advanced Functional Materials (Grant No.MatLiU 2009-00971 through Sweden's innovation agency VINNOVA). I.A.A. is grateful for the Grant from the Ministry of Education and Science of the Russian Federation (Grant No. 14.Y26.31.0005) and Tomsk State University Academic D. I. Mendeleev Fund Program.

**References**

1. C. Freysoldt, B. Grabowski, T. Hickel, J. Neugebauer, G. Kresse, A. Janotti, and C. G. Van de Walle, Rev. Mod. Phys. **86**, 53 (2014).
2. F. Tuomisto and I. Makkonen, Rev. Mod. Phys. **85**, 1583 (2013).
3. G. H. Vineyard, J. Phys. Chem. Solids **3**, 121 (1957).
4. M. Mantina, Y. Wang, R. Arroyave, L. Q. Chen, Z. K. Liu, and C. Wolverton, Phys. Rev. Lett. **100**, 215901 (2008).
5. N. Sandberg, B. Magyari-Köpe, and T. R. Mattsson, Phys. Rev. Lett. **89**, 065901 (2002).
6. O. Hellman, P. Steneteg, I. A. Abrikosov, and S. I. Simak, Phys. Rev. B **87**, 104111 (2013).
7. A. Glensk, B. Grabowski, T. Hickel, and J. Neugebauer, Phys. Rev. X **4**, 011018 (2014).
8. H. M. Gilder and D. Lazarus, Phys. Rev. B **11**, 4916 (1975).
9. O. Hellman, I. A. Abrikosov, and S. I. Simak, Phys. Rev. B **84**, 80301(R) (2011).
10. A.B. Mei, O. Hellman, N. Wireklint, C.M. Schlepütz, D.G. Sangiovanni, B. Alling, A. Rockett, L. Hultman, I. Petrov, J.E. Greene, Phys. Rev. B **91**, 054101 (2015).
11. D. G. Sangiovanni, B. Alling, P. Steneteg, L. Hultman, and I. A. Abrikosov, Phys. Rev. B **91**, 054301 (2015).
12. D. G. Sangiovanni, D. Edström, L. Hultman, I. Petrov, J. E. Greene, and V. Chirita, Surf. Sci. **624**, 25 (2014).
13. D. J. Evans and G. Morriss, (Cambridge University Press, 2008).
14. P. C. Aeberhard, S. R. Williams, D. J. Evans, K. Refson, and W. I. F. David, Phys. Rev. Lett. **108**, 095901 (2012).
15. A. D. Mulliner, P. C. Aeberhard, P. D. Battle, W. I. F. David, and K. Refson, Phys. Chem. Chem. Phys. **17**, 21470 (2015).
16. C. P. Dettmann and G. P. Morriss, Phys. Rev. E **54**, 2495 (1996).
17. S. R. Williams and D. J. Evans, Phys. Rev. Lett. **96**, 015701 (2006).
18. D. J. Evans, W. G. Hoover, B. H. Failor, B. Moran, and A. J. C. Ladd, Phys. Rev. A **28**, 1016 (1983).
19. A. Janotti and C. G. Van de Walle, Phys. Rev. B **76**, 165202 (2007).
20. M. Posselt, F. Gao, and H. Bracht, Phys. Rev. B **78**, 035208 (2008).
21. Q. Shi, J. Voss, H. S. Jacobsen, K. Lefmann, M. Zamponi, and T. Vegge, J. Alloy Compd. **446**, 469 (2007).
22. D. E. Jiang and E. A. Carter, Phys. Rev. B **70**, 064102 (2004).
23. T. Olewicz, G. Antczak, L. Jurczyszyn, J. W. Lyding, and G. Ehrlich, Phys. Rev. B **89**, 235408 (2014).
24. L. J. Xu, G. Henkelman, C. T. Campbell, and H. Jonsson, Surf. Sci. **600**, 1351 (2006).
25. G. Kresse and J. Hafner, Phys. Rev. B **47**, 558 (1993).
26. P. E. Blöchl, Phys. Rev. B **50**, 17953 (1994).
27. R. Armiento and A. E. Mattsson, Phys. Rev. B **72**, 085108 (2005).
28. K. Wang and R. R. Reeber, Mater. Sci. Eng. R **23**, 101 (1998).
29. A. F. Guillermet, Int. J. Thermophys. **6**, 367 (1985).
30. T. R. Mattsson, N. Sandberg, R. Armiento, and A. E. Mattsson, Phys. Rev. B **80**, 224104 (2009).
31. *See supplementary material*.
32. M. Suezawa and H. Kimura, Philos. Mag. **28**, 901 (1973).
33. G. Neumann and V. Tolle, Philos. Mag. **61**, 563 (1990).
34. *See also supplementary material for nearest-neighbor and next nearest-neighbor vacancy jump activation energies calculated at 0 K with the nudged elastic band method*.
35. G. Mills, H. Jonsson, and G. K. Schenter, Surf. Sci. **324**, 305 (1995).
36. K. Maier, H. Mehrer, and G. Rein, Z. Metallkd. **70**, 271 (1979).
37. K. Compaan and Y. Haven, Trans. Faraday Soc. **52**, 786 (1956).

**Figures**

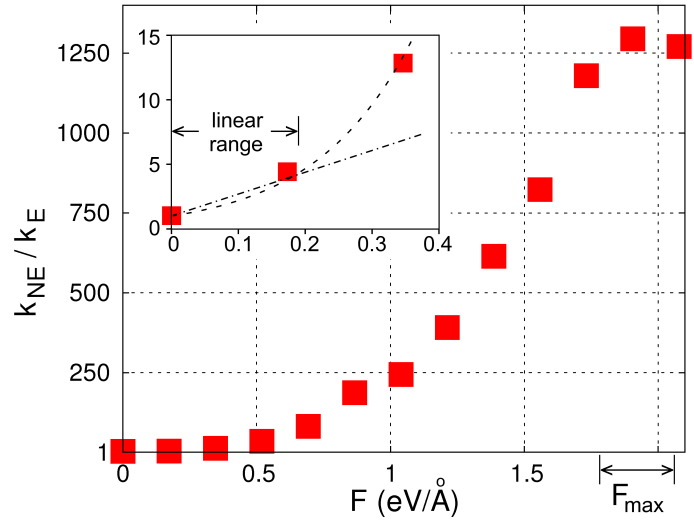

FIG. 1. (Color online). Ratio of the NE-AIMD to AIMD jump rate $k_{NE}/k_E$ obtained as a function of F at 2000 K. The inset illustrates the linear-fitting range, previously used to extrapolate equilibrium rates.[14]

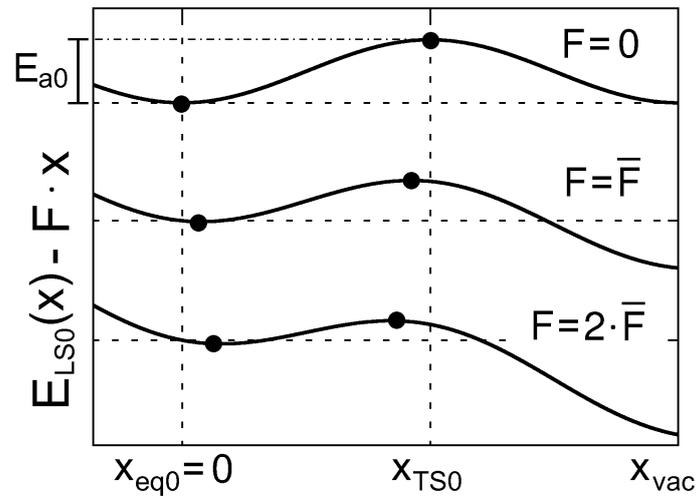

FIG. 2. Model used to clarify the colored-atom jump rate dependence on the intensity F of an applied force seen in Fig. 1. The unperturbed (F = 0) potential energy profile $E_{LS0}$ along the diffusion path x is approximated by a sinusoidal curve. With increasing force intensities F, the colored-atom equilibrium position $x_{eq}$ and the transition-state coordinate $x_{TS}$ move toward each other, while the effective jump activation energy $E_a$ decreases monotonically. $x_{eq0}$, $x_{TS0}$, and $E_{a0}$ are zero-field quantities. $x_{vac}$ indicates the vacancy position.

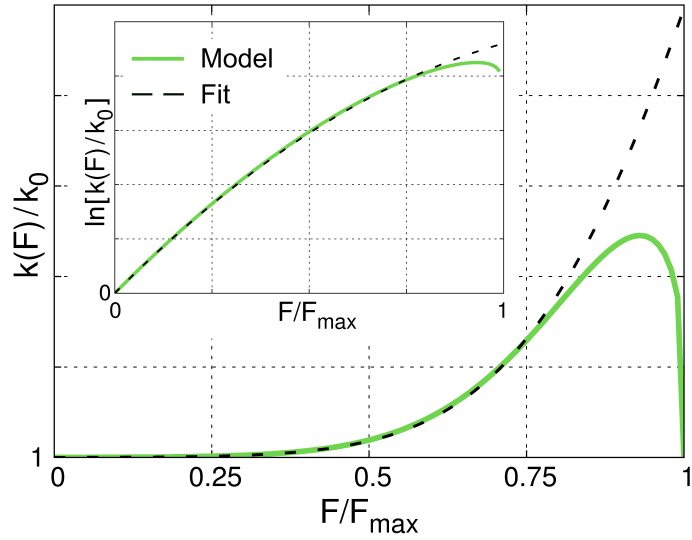

FIG. 3. (Color online). Numerical evaluation of accelerated-rates k(F) as a function of the intensity F of the force applied to the colored-atom based on the model illustrated in Fig. 2, and fit of k(F) up to $F' \simeq 0.75\, F_{max}$ [Eqs. (1) and (2)].

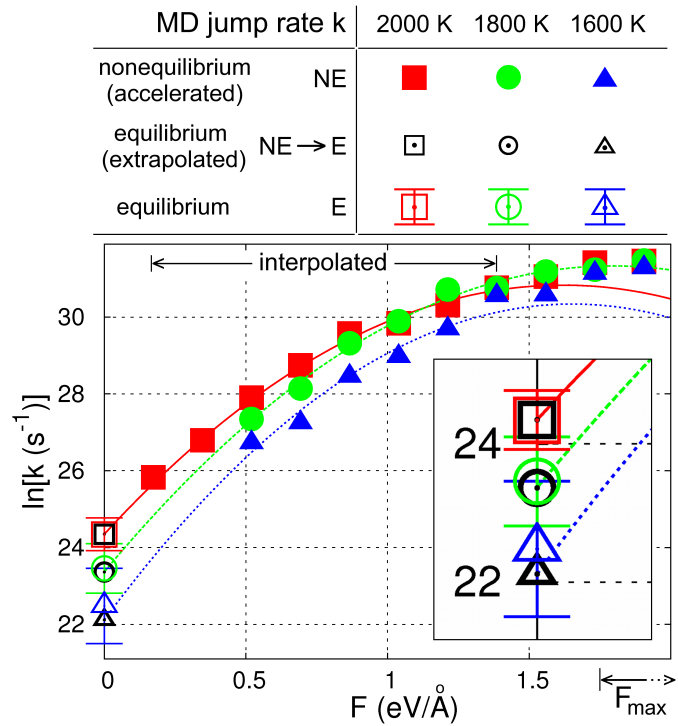

FIG. 4. (Color online). Nonequilibrium jump rates $\ln[k_{NE}(F,T)]$ obtained as a function of F for T = 1600, 1800, and 2000 K. The inset illustrates the agreement between equilibrium rates $k_{NE \to E}$ extrapolated by nonequilibrium $k_{NE}$ results and equilibrium rates $k_E$ obtained by non-accelerated AIMD. $\ln(k_E)$ is shown with corresponding error bars. The force limit used for $k_{NE}$ result fitting, as determined in our model calculations (Fig. 3), is $F' \simeq 0.75\, F_{max}$.

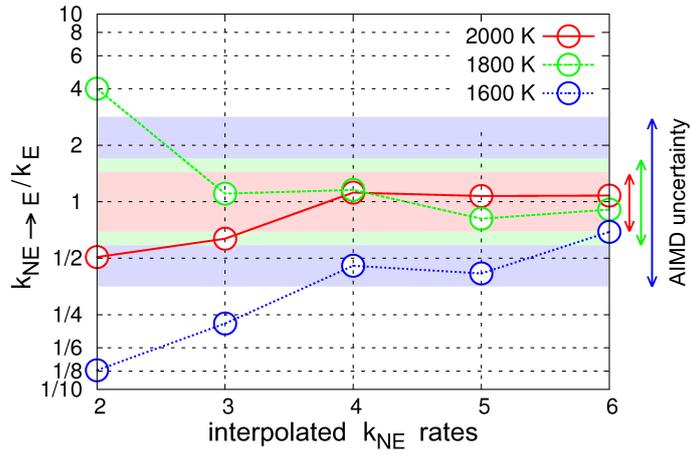

FIG. 5. (Color online). Convergence of extrapolated equilibrium rates $k_{NE \to E}$, normalized by AIMD rate $k_E$ values, with the number of nonequilibrium $k_{NE}$ interpolation points. $k_{NE \to E}/k_E$ ratios are obtained for f = 0.7 and 0.8 eV/Å (two interpolated rates), and f ranging between 0.3 and 0.8 eV/Å (six interpolated rates). The vertical scale is logarithmic (basis 10). Shaded red, green, and blue regions with limits marked by arrows (on the right) correspond to $k_E$ error bars at 2000, 1800, and 1600 K, respectively.

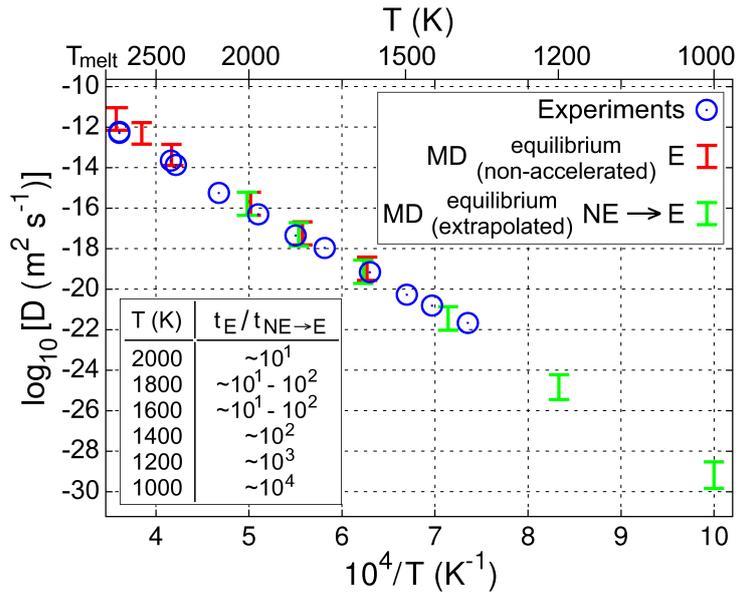

FIG. 6. (Color online). Comparison between theoretical and experimental diffusion coefficients D.[36] The error bars on $D_E(T)$ ($\propto k_E(T) \cdot c_V(T)$) and $D_{NE \to E}(T)$ ($\propto k_{NE \to E}(T) \cdot c_V(T)$) values account only for the experimental uncertainty on equilibrium vacancy concentrations $c_V(T)$. The table in the inset lists the gain in computational efficiency $t_E/t_{NE \to E}$ achieved by our approach, where $t_E$ and $t_{NE \to E}$ are approximate simulation times required to obtain well-converged equilibrium vacancy jump rates as a function of T in non-accelerated AIMD and accelerated NE-AIMD simulations, respectively.

**Supplemental Material**

**I.(S) Derivation of NE-AIMD rate k(F, T) dependence on F**

The unperturbed effective potential energy profile $E_{LS0}$ (along straight path x connecting the colored-atom to the vacancy site) is approximated as:

$$E_{LS0}(x) = \frac{E_{a0}^{model}}{2}\left[1 - \cos(\pi \cdot x / x_{TS0})\right]. \qquad (S.1)$$

A thermally-equilibrated system is immersed in a constant force-field **F** directed along x. The force acting on the colored-atom has intensity F. After a short transient period, the system reaches a nonequilibrium steady-state, and the effective potential energy profile becomes (Fig. 1.S):

$$E_{LS}(x, F) = E_{LS0}(x) - F \cdot x = \frac{E_{a0}}{2}\left[1 - \cos(\pi \cdot x / x_{TS0})\right] - F \cdot x. \qquad (S.2)$$

$E_{LS}$ first derivative in x:

$$\frac{\partial E_{LS}(x, F)}{\partial x} = \frac{\pi \cdot E_{a0}}{2\, x_{TS0}} \cdot \sin(\pi \cdot x / x_{TS0}) - F. \qquad (S.3)$$

$E_{LS}$ and $E_{LS0}$ second derivatives in x are equivalent:

$$\frac{\partial^2 E_{LS}(x, F)}{\partial x^2} = \frac{\partial^2 E_{LS0}(x)}{\partial x^2} = \frac{\pi^2 \cdot E_{a0}}{2\, x_{TS0}^2} \cdot \cos(\pi \cdot x / x_{TS0}). \qquad (S.4)$$

For increasing force-field intensities F, the colored-atom equilibrium position $x_{eq}$ and the transition state $x_{TS}$ shift continuously toward each other (Fig. 1.S):

$$\frac{\partial E_{LS}(x, F)}{\partial x} = 0 \;\Rightarrow\; x = \begin{cases} \dfrac{x_{TS0}}{\pi} \cdot \left\{\arcsin\left[\dfrac{2\, F \cdot x_{TS0}}{\pi \cdot E_{a0}}\right] + 2n \cdot \pi\right\}; \forall n \in \mathbb{Z} \\[2ex] \dfrac{x_{TS0}}{\pi} \cdot \left\{-\arcsin\left[\dfrac{2\, F \cdot x_{TS0}}{\pi \cdot E_{a0}}\right] + (2n+1) \cdot \pi\right\}; \forall n \in \mathbb{Z} \end{cases} \;\Rightarrow$$

$$x_{eq}(F) = \frac{x_{TS0}}{\pi} \cdot \arcsin\left[\frac{2\,F \cdot x_{TS0}}{\pi \cdot E_{a0}}\right] \quad , \quad x_{TS}(F) = \frac{x_{TS0}}{\pi} \cdot \left\{\pi - \arcsin\left[\frac{2\,F \cdot x_{TS0}}{\pi \cdot E_{a0}}\right]\right\}. \tag{S.5}$$

The distance between the equilibrium position $x_{eq}(F)$ and the transition state $x_{TS}(F)$ varies as a function of F as:

$$x_{TS}(F) - x_{eq}(F) = \frac{x_{TS0}}{\pi} \cdot \left\{\pi - 2 \cdot \arcsin\left[\frac{2\,F \cdot x_{TS0}}{\pi \cdot E_{a0}}\right]\right\}. \tag{S.6}$$

The intensity of the force field used in NE-AIMD runs must be lower than the limit $F_{max}$, for which $x_{eq}(F) = x_{TS}(F)$:

$$x_{TS}(F) - x_{eq}(F) = 0 \;\Rightarrow\; \arcsin\left[\frac{2\,F \cdot x_{TS0}}{\pi \cdot E_{a0}}\right] = \frac{\pi}{2} \;\Rightarrow\; \frac{2\,F \cdot x_{TS0}}{\pi \cdot E_{a0}} = 1 \;\Rightarrow\; F_{max} = \frac{\pi \cdot E_{a0}}{2\,x_{TS0}}. \tag{S.7}$$

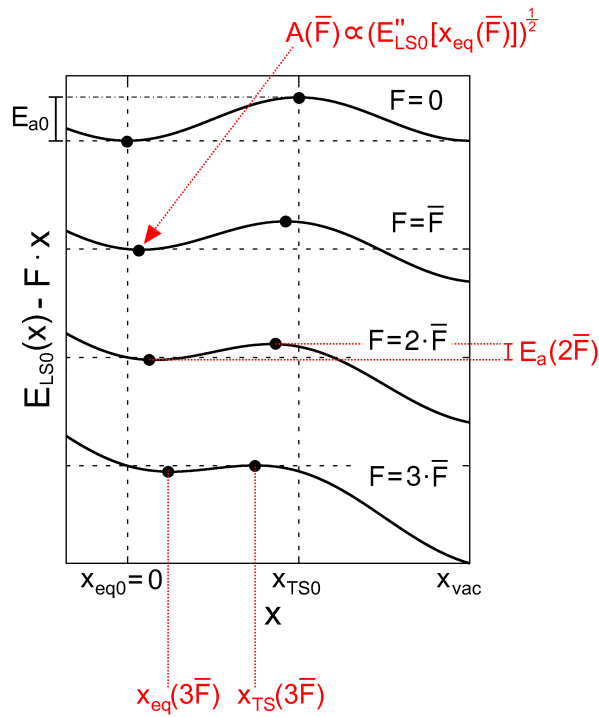

Figure 1.S. Schematic illustration of the effects of increasing F on $E_a(F)$, $A(F)$, $x_{eq}(F)$, and $x_{TS}(F)$.

The jump activation energy $E_a$ varies as a function of F as (see Figs. 1.S and 2.S):

$$E_a(F) = E_{LS}\left[x_{TS}(F), F\right] - E_{LS}\left[x_{eq}(F), F\right] =$$

$$= E_{a0} \cdot \sqrt{1 - \left[\frac{2\,F \cdot x_{TS0}}{\pi \cdot E_{a0}}\right]^2} - F \cdot x_{TS0} + 2\,F \cdot \frac{x_{TS0}}{\pi} \cdot \arcsin\left[\frac{2\,F \cdot x_{TS0}}{\pi \cdot E_{a0}}\right] \Rightarrow$$

$$E_a(F) = E_{a0} \cdot \sqrt{1 - \left(\frac{F}{F_{max}}\right)^2} + F \cdot x_{TS0}\left[\frac{2}{\pi} \cdot \arcsin\left(\frac{F}{F_{max}}\right) - 1\right]. \tag{S.8}$$

The jump attempt frequency varies as a function of F as (see Figs. 1.S and 2.S):

$$A(F) = A_0 \cdot \frac{\sqrt{\left[\partial^2 E_{LS0}(x)/\partial^2 x\right]_{x_{eq}(F)}}}{\sqrt{\left[\partial^2 E_{LS0}(x)/\partial^2 x\right]_{x_{eq}=0}}} = A_0 \cdot \sqrt{\frac{x_{TS0}}{\pi \cdot F_{max}}} \cdot \sqrt{\left[\partial^2 E_{LS}(x, F)/\partial^2 x\right]_{x_{eq}(F)}} =$$

$$= A_0 \cdot \sqrt{\cos\left(\arcsin\left[\frac{2\,F \cdot x_{TS0}}{\pi \cdot E_{a0}}\right]\right)} = A_0 \cdot \left[1 - \left(\frac{F}{F_{max}}\right)^2\right]^{\frac{1}{4}} \Rightarrow A(F) = A_0 \cdot \left[1 - \left(\frac{F}{F_{max}}\right)^2\right]^{\frac{1}{4}}. \tag{S.9}$$

Introducing a normalized force $F_n = F/F_{max}$ ranging between 0 and 1, Eqs. (S.8) and (S.9) become:

$$E_a(F_n) = E_{a0} \cdot \left\{\sqrt{1 - F_n^2} + F_n \cdot \left[\arcsin(F_n) - \frac{\pi}{2}\right]\right\}, \tag{S.10}$$

$$A(F_n) = A_0 \cdot \left[1 - F_n^2\right]^{\frac{1}{4}}. \tag{S.11}$$

Taylor expansion of $E_a$ and A up to the 4$^{th}$ order in $F_n$ yields:

$$E_a(F_n) = E_{a0}\left[1 - \frac{\pi \cdot F_n}{2} + \frac{F_n^2}{2} + \frac{F_n^4}{24}\right] + O(F_n^5) \cong E_{a0}\left[1 - \frac{\pi \cdot F_n}{2} + \mu \cdot F_n^2\right] \quad \text{with} \quad \mu > 0, \tag{S.12}$$

$$A(F_n) = A_0 \cdot \left(1 - \frac{F_n^2}{4} + \frac{F_n^4}{32}\right) + O(F_n^5) = A_0 \cdot \left[\exp\left(-\frac{F_n^2}{4}\right) - \frac{F_n^4}{8}\right] + O(F_n^5) \Rightarrow$$

$$A(F_n) \cong A_0 \cdot \exp(-\phi \cdot F_n^2) \quad \text{with} \quad \phi > 0. \tag{S.13}$$

µ and ϕ are fitting parameters.

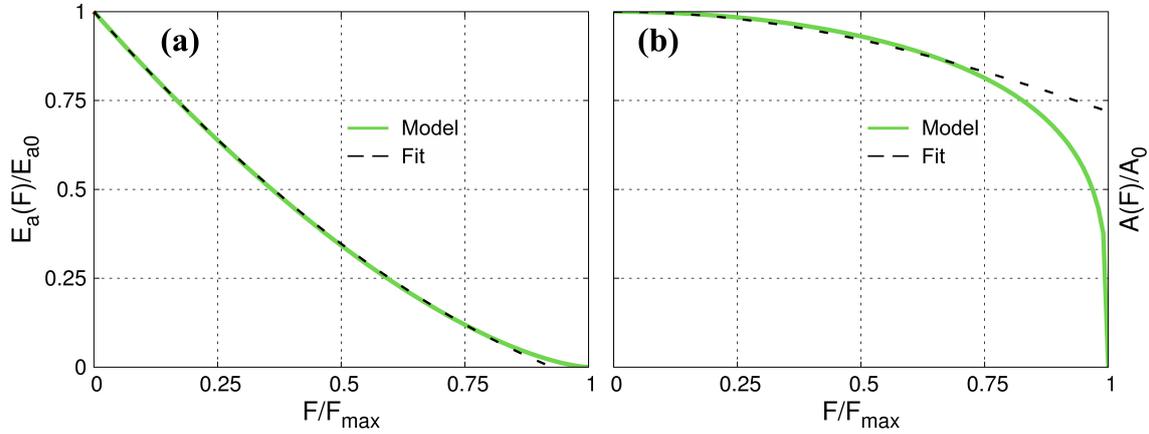

Figure 2.S. (a) $E_a(F)/E_{a0}$ and (b) $A(F)/A_0$ dependence on F as obtained from Eqs. (S.10) and (S.11) (labeled as *Model*) vs. fitted curves (Eqs. S.12 and S.13) with parameters µ = 0.53, and ϕ = 0.33 (labeled as *Fit*).

Using $A(F_n)$ and $E_a(F_n)$ [Eqs. (S.12) and (S.13)] in the Arrhenius relation, the jump rate k at a given temperature T varies as a function of F as:

$$k(F_n, T) = A(F_n) \cdot \exp[-E_a(F_n)/(k_B T)] = A_0 \cdot \exp(-\phi \cdot F_n^2) \cdot \exp\left(\frac{-E_{a0}\left[1 - \frac{\pi \cdot F_n}{2} + \mu \cdot F_n^2\right]}{k_B T}\right) =$$

$$= k_0(T) \cdot \exp\left(\frac{E_{a0}\left[\frac{\pi \cdot F_n}{2} - \mu \cdot F_n^2\right]}{k_B T} - \phi \cdot F_n^2\right) \Rightarrow k(F_n, T) = k_0(T) \cdot \exp\left[\frac{\pi \cdot E_{a0}}{2 k_B T} \cdot F_n - \left(\frac{\mu \cdot E_{a0}}{k_B T} + \phi\right) \cdot F_n^2\right].$$

The jump activation energy $E_{a0}$ varies slowly with T. In the present analysis we treat it as a constant. Substituting back $F_n$ with $F/F_{max}$, the previous expression becomes:

$$k(F,T) = k_0(T) \cdot \exp\left[\frac{x_{TS0}(T)}{k_B T} \cdot F - \left(\frac{\mu \cdot E_{a0}}{k_B T} + \phi\right) \cdot \frac{1}{F_{max}^2} \cdot F^2\right] \Rightarrow$$

$$k(F,T) = k_0(T) \cdot \exp\left[\frac{x_{TS0}(T)}{k_B T} \cdot F - \alpha(T) \cdot F^2\right], \tag{S.14}$$

for which:

$$\alpha(T) = \frac{1}{F_{max}^2} \cdot \left(\frac{\mu \cdot E_{a0}}{k_B T} + \phi\right). \tag{S.15}$$

Eq. (S.14) corresponds to Eq. (1) in the paper.

**II.(S) 0 K potential energy profiles for nearest-neighbor and next nearest-neighbor Mo jumps**

The potential energy profile along the minimum-energy migration path of Mo atoms into a neighboring vacancy site are estimated with the nudged elastic band (NEB) method [35] as implemented in VASP. The ionic positions are relaxed by DFT+AM05 calculations employing 6x6x6 k-point grids and a cutoff of 500 eV for the plane-wave basis set until Hellman-Feynman forces and total energy converge within $10^{-2}$ eV/Å and $10^{-5}$ eV/atom, respectively. NEB calculations employ the VASP default spring parameter (SPRING = -5) and 11 images between the initial and final state. For nearest-neighbor jumps, an additional NEB run is carried out to obtain a thicker set of data in the vicinity of the transition state.

NEB yields activation energies of 1.29 eV, in agreement with previous *ab initio* estimations [30], for nearest-neighbor (NN) jump and 4.09 eV for next-nearest neighbor (NNN) jumps (Fig. 3.S). Both minimum-energy paths are straight and correspond to migration distances of 2.650 Å along <111> directions for NN and 3.040 Å along <100> directions for NNN jumps. At 0 K, the energy-barrier for NNN vacancy migration is considerably larger than that of NN migration. As NNN vacancy migration is not detected in AIMD runs carried out at temperatures close to melting, it can be concluded that NNN jumps yield negligible contribution to monovacancy diffusion in bcc Mo at all temperatures.

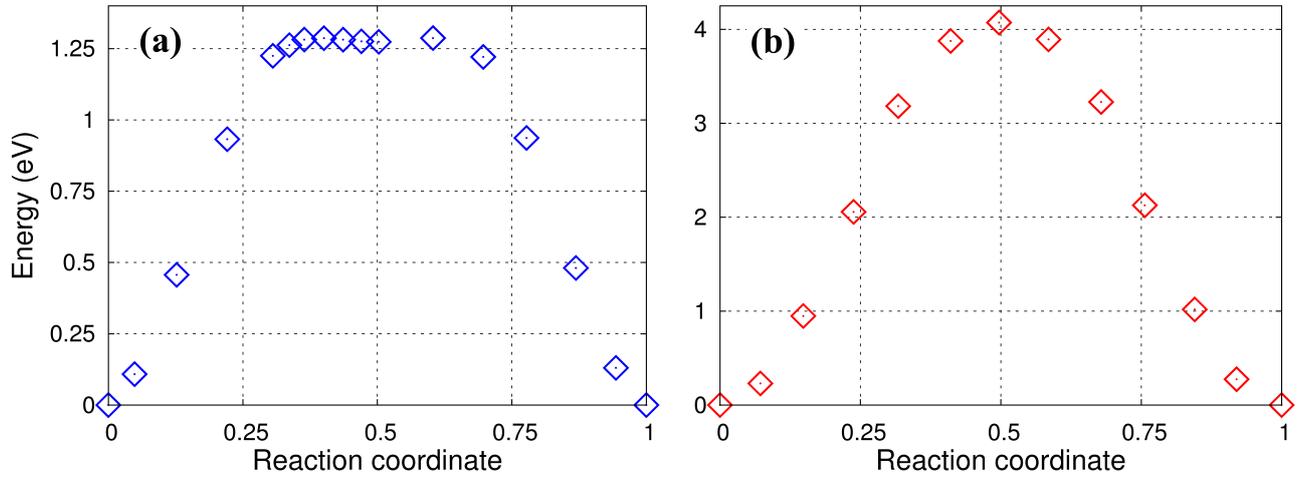

Figure 3.S. DFT+NEB 0 K potential energy profile for (a) nearest-neighbor and (b) next nearest-neighbor vacancy jumps in bcc Mo.

**III.(S) Identification of nearest-neighbor and next nearest-neighbor lattice-atom jumps**

Nearest neighbor (NN) and next nearest-neighbor (NNN) lattice-atom jumps are detected using a script which parses the MD atomic coordinates at each simulation time step. The reliability of the results obtained with this script is double-checked by careful comparison with plots of NN and NNN positions as a function of time (vacancy jump corresponds to a sudden change in atomic coordinates), as well as by directly watching MD movies for several ps.

Prior to recording migration events in MD simulations, the system is equilibrated at a given temperature for a few ps. Thus, the MD time $t$ is set to zero and the reference coordinate $\vec{r}_{static}(j)$ of a bcc lattice-site $j$ ($j = 1, …, n$) in a static defect-free crystal (containing $n$ atoms) is associated to the closest (smallest difference in fractional coordinates) atom $i$ at position $\vec{r}_{MD}(i, t=0)$ of the defective structure. The reference coordinate $\vec{r}_{static}(j_V)$ for which the distance

$$\delta(i, j_V, t=0) = \left|\vec{r}_{static}(j_V) - \vec{r}_{MD}(i, t=0)\right| \geq d_{NN} - \varepsilon \quad \forall \ i \in [1:n-1] \qquad (S.16)$$

(the tolerance $\varepsilon$ is small compared to the nearest-neighbor distance $d_{NN}$) is assigned to the lattice vacancy in the defective structure. For cases in which at $t = 0$ the condition (S.16) was not verified for all $n$-1 lattice atoms $i$, system equilibration would be extended for an additional 0.1 ps.

For any MD time-step $t = t' > 0$, NN and NNN jumps are detected using the conditions:

$$\text{NN jump} := \begin{cases} \delta(i, j_V, t') < \varepsilon \\ \delta(i, j_V, t') \geq d_{NN} - \varepsilon \\ \delta(i, j_V, t') \leq d_{NN} + \varepsilon \end{cases} \quad ; \quad \text{NNN jump} := \begin{cases} \delta(i, j_V, t') < \varepsilon \\ \delta(i, j_V, t') \geq \frac{2 \cdot d_{NN}}{\sqrt{3}} - \varepsilon \\ \delta(i, j_V, t') \leq \frac{2 \cdot d_{NN}}{\sqrt{3}} + \varepsilon \end{cases} . \quad (S.17)$$

When, at a MD time $t = \bar{t}$, one of the sets of requirements (S.17) is met (a lattice-atom $i_{\text{jump}}$ has migrated into the vacancy), the script swaps the two reference coordinates: $\vec{r}_{\text{static}}(j_V)$ is associated to atom $i_{\text{jump}}$, while $\vec{r}_{\text{static}}(j_{\text{jump}})$ is assigned to the newly formed lattice vacancy.